\documentclass[pra,aps,preprint,showpacs]{revtex4}
\usepackage{bm}
\usepackage{amsmath,amssymb}
\usepackage{graphicx}

\begin{document}

\title{Multiphoton ionization of xenon in the vuv regime}
\author{Robin Santra}
\author{Chris H. Greene}
\affiliation{Department of Physics and JILA, University of Colorado,
Boulder, CO 80309-0440, USA}
\date{\today}
\begin{abstract}
In a recent experiment at the vuv free-electron laser facility at DESY in Hamburg, the
generation of multiply charged ions in a gas of atomic xenon was observed. This paper
develops a theoretical description of multiphoton ionization of xenon and its ions.
The numerical results lend support to the view that the experimental observation may 
be interpreted in terms of the nonlinear absorption of several vuv photons. The method
rests on the Hartree-Fock-Slater independent-particle model. The multiphoton physics is
treated within a Floquet scheme. The continuum problem of the photoelectron is solved
using a complex absorbing potential. Rate equations for the ionic populations are integrated
to take into account the temporal structure of the individual vuv laser pulses. The effect
of the spatial profile of the free-electron laser beam on the distribution of xenon charge
states is included. An Auger-type many-electron mechanism may play a role in the vuv
multiphoton ionization of xenon ions.
\end{abstract}
\maketitle

\section{Introduction}
\label{sec1}

An electron bound to an atom experiences electric forces, which on average point
toward the atomic nucleus. If the atom is placed in a static electric field, the
electronic states become unstable, as the potential arising from the superposition
of the atomic and the external electric field enables electron emission via tunneling. 
This picture basically remains valid even if the external electric field is oscillating, 
at least as long as the oscillation period of the electric field is long in comparison 
to the electron tunneling timescale \cite{Keld64,Fais73,Reis80,AmDe86,Cork93}.
For a typical valence-electron binding energy of the order of $10$ eV and a laser photon 
energy of about $1$ eV---i.e., in the ir regime---the tunneling picture is meaningful 
for intensities of $10^{13}$ W/cm$^2$ or higher. However, at high radiation frequency (or 
at low intensity), this picture fails, and it is more appropriate 
to adopt a multiphoton picture \cite{EbLa78,Fais87,SmKn88,Mitt93}: An atomic electron can 
be ejected following the absorption of a discrete number of photons.

The development of free-electron laser (FEL) \cite{FEL99} facilities at several places in the world
\cite{FEL03} has spurred substantial theoretical interest 
\cite{BrRz99,GaKr00,NoHo00,BaPl01,IsMi02,NoHo02,SaRo02,NiNa03,BaRo03,SaCe03}, but until very 
recently, no radiation sources have been available to experimentally probe strong-field 
physics at vuv or shorter wavelengths. The situation changed when the VUV-FEL at DESY, Hamburg,
began operation \cite{AnAu00,AyBa02}. In one of the first experiments with this exceptional
radiation source, xenon clusters were exposed to the intense vuv laser pulses and were observed
to absorb a surprisingly large number of vuv photons per atom \cite{WaBi02}, a finding which was
explained in terms of inverse bremsstrahlung \cite{SaGr03}. No evidence for atomic multiphoton 
processes was found in these first measurements. Measurements using a more sensitive detector, 
however, revealed the creation of multiply charged ions even in a gas of free, unclustered xenon
atoms \cite{WaCa04}. An experimental time-of-flight mass spectrum, averaged over $100$ consecutive
VUV-FEL pulses, is shown in Fig. \ref{fig1} \cite{MoWa04}. The photon energy in that experiment 
was $12.7$ eV with a peak intensity of approximately $10^{13}$ W/cm$^2$.

In this paper, we present a theoretical description of the interaction of xenon atoms with an
intense pulse of vuv photons. Our findings lend support to the interpretation that the occurrence
of multiply charged ions, as reported in Ref.~\cite{WaCa04}, is a consequence of vuv multiphoton
ionization---a scientific first at a photon energy of more than $10$ eV. We describe our 
computational method in Sec.~\ref{sec2}. In Sec.~\ref{sec3}, the results of our calculations are 
presented and compared with experiment. We conclude with Sec.~\ref{sec4}. Atomic units are used
throughout, unless otherwise noted.

\section{Computational Method}
\label{sec2}

Using an independent-electron model, Geltman \cite{Gelt85} (see also Refs. \cite{Zakr86,GeZa88}) 
was able to arrive at a satisfactory representation of the ionic charge distributions observed in
several intense-laser experiments at photon energies of $6.4$ eV and below 
\cite{HuLo82,LuPu83,HuLI83,HuII83,YeCh87}. Within the framework of independent particles,
each electron moves in the field of the atomic nucleus and in a mean-field generated by the other
electrons. The best such mean-field derives from the Hartree-Fock self-consistent-field method
\cite{SzOs96}. However, the Hartree-Fock mean-field is nonlocal, due to the exchange interaction,
and therefore cumbersome to work with. Slater \cite{Slat51} introduced a local approximation to
electron exchange, which is the principle underlying the well-known X$\alpha$ method \cite{SlJo72}.

In this work, we use the Hartree-Fock-Slater code written by Herman and Skillman \cite{HeSk63}, which 
in the past has proved advantageous for atomic photoionization studies (see, for example, Ref. \cite{MaCo68}).
The resulting one-electron potential, $V_{\mathrm{HS}}(r)$, is a central potential (even for
open-shell systems), which satisfies
\begin{eqnarray}
\label{eq1}
V_{\mathrm{HS}}(r) \rightarrow -Z/r & , & r \rightarrow 0 \\
\label{eq2}
V_{\mathrm{HS}}(r) \rightarrow -(q+1)/r & , & r \rightarrow \infty 
\end{eqnarray}
for an atom of nuclear charge $+Z$ and overall charge $+q$. In the original program of Herman and
Skillman, the X$\alpha$ parameter was set to unity, in accordance with Ref. \cite{Slat51}. We
adjusted that parameter slightly (see Table \ref{tab1}), so that the calculated $5p$ binding energy
in Xe, Xe$^+$, Xe$^{++}$, and Xe$^{3+}$ agrees with experimental data \cite{BrVe01,HaPe87,MaBa87,GrDi83}.
The situation is a little more complex for Xe$^{4+}$ and Xe$^{5+}$. The experimental ionization
potentials of these ions, as determined from measured electron-impact ionization thresholds
\cite{DoMo59,Stub65,Syag92}, vary by as much as $18$ eV. Utilizing the complete-active-space
self-consistent-field (CASSCF) code implemented in the {\em ab initio} program package MOLPRO
\cite{WeKn85,KnWe85}; an active space consisting of the $4d$, $5s$, $5p$, $6s$, and $6p$ orbitals;
and an effective core potential \cite{JoCh87}, we obtained ionization potentials of $53$ and $64$ eV
for Xe$^{4+}$ and Xe$^{5+}$, respectively. The value for Xe$^{4+}$ agrees with Ref. \cite{DoMo59};
the one for Xe$^{5+}$ is in agreement with Ref. \cite{Syag92}. As indicated in Table \ref{tab1}, we base
the X$\alpha$ parameters for Xe$^{4+}$ and Xe$^{5+}$ on the ionization potentials quoted above. 
It should be mentioned that while our CASSCF calculation includes scalar relativistic effects in an 
approximate manner (through the effective core potential), spin-orbit splittings have not been taken 
into account. 

We represent the radial one-electron Hamiltonian 
\begin{equation}
\label{eq3}
H^{(l)}_{\mathrm{AT}} = -\frac{1}{2}\frac{\mathrm{d}^2}{\mathrm{d}r^2} + \frac{l(l+1)}{2r^2} 
+ V_{\mathrm{HS}}(r)
\end{equation}
in a finite-element basis set \cite{Bath76,BaWi76,BrSc93,AcSh96,ReBa97,MeGr97}, which is described in 
detail in Ref. \cite{SaCh04}. In the calculations reported here, $4801$ finite-element basis functions 
were employed, spanning a radial grid from $r_{\mathrm{min}} = 0$ to $r_{\mathrm{max}} = 80$ bohr radii.
For each orbital angular momentum quantum number, $l$, considered, the first $50$ eigenfunctions 
$u_{n,l}(r)$ of $H^{(l)}_{\mathrm{AT}}$ were computed that have eigenenergies $\varepsilon_{n,l}$ at 
least as high as the energy $\varepsilon_{5,0}$ of the $5s$ level. More strongly bound levels are 
assumed to be fully occupied by core electrons and are not considered any further. The calculated 
eigenfunctions $u_{n,l}(r)$ satisfy the boundary conditions $u_{n,l}(r_{\mathrm{min}}) = 0$ and 
$u_{n,l}(r_{\mathrm{max}}) = 0$. At the current level of approximation, atomic multiplet structure is 
absent. All electrons that can be associated with the quantum numbers $n$ and $l$ have the same energy, 
$\varepsilon_{n,l}$, within our model. In particular, there is no energy dependence on the magnetic 
quantum number, $m$. 

In order to treat the problem of electron emission, a complex absorbing potential (CAP) 
\cite{GoSh78,JoAu85,JoAu86,NeBa89,RiMe93,RiMe95,Mois98,RiMe98,PaMu98,PaMuII,Karl98,SoRi98,SaCe01,SaCe02,Mano02,PoCa03,PoCaII},  
$-{\mathrm{i}} \eta W(r)$, is added to the atomic Hamiltonian. The real, positive parameter $\eta$ is the
CAP strength. The local one-electron potential $W(r)$ is chosen here as 
\begin{eqnarray}
\label{eq4}
W(r) & = & \left\{
            \begin{array}{cl}
            0, &\: 0 \leq r < c \\
            (r-c)^2, &\: r \geq c
            \end{array}
            \right. , 
\end{eqnarray}
where $c = 4$ a.u. in this paper (which places the absorbing potential right outside the ionic core). 
The CAP absorbs the outgoing electron and renders the associated wave function square-integrable. 
Given a complete basis, there exists for a resonance eigenstate (a Gamow vector) of the physical 
Hamiltonian with Siegert eigenvalue $E_{\mathrm{res}} = E_{\mathrm{R}} - {\mathrm{i}} \Gamma/2$, 
an eigenvalue $E(\eta)$ of the CAP-augmented Hamiltonian such that 
$\lim_{\eta \to 0^+} E(\eta) = E_{\mathrm{res}}$ \cite{RiMe93}. In a finite basis set, 
$\eta_{\mathrm{opt}}$ must be found, satisfying \cite{RiMe93}
\begin{equation}
  \label{eq5}
  \left| \eta \frac{{\mathrm{d}} E(\eta)}{{\mathrm{d}} \eta} \right|_{\eta_{\mathrm{opt}}}
  = \hbox{minimum} \; .
\end{equation}
$E(\eta_{\mathrm{opt}})$ provides an approximation to the Siegert energy $E_{\mathrm{res}}$,
from which the resonance position, $E_{\mathrm{R}}$, and the resonance width, $\Gamma$, can be 
extracted. An improved strategy, which is used here, consists in analyzing the function
$E(\eta) - \eta {\mathrm{d}} E(\eta) / {\mathrm{d}} \eta$ and minimizing 
$\left| \eta^{2} {\mathrm{d}}^{2} E(\eta)/ {\mathrm{d}} \eta^{2} \right|$ \cite{RiMe93}.

Within the framework of quantum electrodynamics \cite{CrTh98}, the Hamiltonian describing the
effective one-electron atom interacting with the electromagnetic field, in the presence of the CAP,
reads 
\begin{equation}
  \label{eq6}
  H = H_{\mathrm{AT}} + H_{\mathrm{EM}} + H_{\mathrm{I}} -{\mathrm{i}} \eta W \; ,
\end{equation}
where
\begin{eqnarray}
  \label{eq7}
  H_{\mathrm{AT}} & = & -\frac{1}{2}\nabla^2 + V_{\mathrm{HS}}(r) \; , \\
  \label{eq8}
  H_{\mathrm{EM}} & = & \sum_{{\bm k},\lambda} \omega a^{\dag}_{{\bm k},\lambda} a_{{\bm k},\lambda} \; , \\
  \label{eq9}
  H_{\mathrm{I}}  & = & {\bm x} \cdot \sum_{{\bm k},\lambda} {\mathrm{i}} \sqrt{\frac{2 \pi}{V}\omega}
  \left\{{\bm e}_{{\bm k},\lambda} a_{{\bm k},\lambda} - 
        {\bm e}^{\ast}_{{\bm k},\lambda} a^{\dag}_{{\bm k},\lambda}\right\} \; .
\end{eqnarray}
Here, $H_{\mathrm{EM}}$ represents the free electromagnetic field, $H_{\mathrm{I}}$ the interaction term in
electric-dipole approximation (in length gauge). The operator $a^{\dag}_{{\bm k},\lambda}$
($a_{{\bm k},\lambda}$) creates (annihilates) a photon with wave vector  ${\bm k}$, polarization 
$\lambda$, and energy $\omega = k/\alpha$ ($\alpha$ is the fine-structure constant). We use the symbol
${\bm x}$ for the atomic dipole operator. $V$ in Eq.~(\ref{eq9}) denotes the normalization volume of the 
electromagnetic field, and ${\bm e}_{{\bm k},\lambda}$ indicates the polarization vector of mode 
${\bm k},\lambda$.

Let $N$ be the number of photons in the laser mode, so that the laser intensity is given by
\begin{equation}
  \label{eq10}
  I = \frac{N}{V}\frac{\omega}{\alpha} \; ,
\end{equation}
intensity being measured in units of $I_0 = E_{\mathrm{h}}/(t_0 a_0^2) = 6.43641 \times 10^{15}$ W/cm$^2$
($E_{\mathrm{h}}$: Hartree energy; $t_0$: atomic unit of time; $a_0$: Bohr radius). We can now combine
the atomic eigenstates, $\psi_{n,l,m} = (u_{n,l}(r)/r) Y_{l,m}(\vartheta,\varphi)$, with the Fock 
states of the laser mode, $\left|N-\mu\right>$ ($\mu = 0, \pm 1, \pm 2, \ldots$), to form basis vectors
$\left|\Phi_{n,l,m,\mu}\right> = \left|\psi_{n,l,m}\right>\left|N-\mu\right>$. Assuming linear 
polarization, the matrix representation of the Hamiltonian $H$ [Eq. (\ref{eq6})] in the basis 
$\left\{\left|\Phi_{n,l,m,\mu}\right>\right\}$ is diagonal with respect to $m$. It also has a rather 
sparse structure with respect to $n$, $l$, and $\mu$. The only nonzero matrix elements are
\begin{widetext}
\begin{eqnarray}
  \label{eq11}
\left<\Phi_{n,l,m,\mu}\right|H_{\mathrm{AT}} + H_{\mathrm{EM}}\left|\Phi_{n,l,m,\mu}\right>
 & = & \varepsilon_{n,l} - \mu\omega \; , \\
  \label{eq12}
\left<\Phi_{n,l,m,\mu}\right|H_{\mathrm{I}}\left|\Phi_{n',l',m,\mu+1}\right>
 & = & \sqrt{2\pi\alpha I} \left<\psi_{n,l,m}\right|z\left|\psi_{n',l',m}\right> \; , \\
  \label{eq13}
\left<\Phi_{n,l,m,\mu+1}\right|H_{\mathrm{I}}\left|\Phi_{n',l',m,\mu}\right>
 & = & \sqrt{2\pi\alpha I} \left<\psi_{n,l,m}\right|z\left|\psi_{n',l',m}\right> \; , \\
  \label{eq14}	
\left<\Phi_{n,l,m,\mu}\right|W\left|\Phi_{n',l,m,\mu}\right> & = & 
\left<\psi_{n,l,m}\right|W\left|\psi_{n',l,m}\right> \; .
\end{eqnarray}
\end{widetext}
The energy $N\omega$ of the unperturbed laser field has been subtracted from the right-hand side of 
Eq. (\ref{eq11}); the relatively high intensity ($N \gg |\mu|$) has been exploited in the coupling 
matrix blocks (Eqs. (\ref{eq12}) and (\ref{eq13})); and a unitary transformation has been applied that
renders the matrix ${\bm H}_{\mathrm{AT}} + {\bm H}_{\mathrm{EM}} + {\bm H}_{\mathrm{I}}$ real 
symmetric. The complete matrix ${\bm H}$ is complex symmetric and of the Floquet-type
\cite{Shir65,ChRh77,ChCo85,BuFr91,DoTe92,ChTe04} (see, for example, 
Refs. \cite{Kula87,Kula88,HuPi97,TaDu99,LaSt02} and references therein for other computational approaches 
to atomic strong-field physics). In our calculations, $\mu$ runs from $\mu_{\mathrm{min}} = 0$ to 
$\mu_{\mathrm{max}}$, the minimum number of photons needed to photoionize (referred to as N.P. in Table \ref{tab1}). 
Thus, since electron emission can only take place after the absorption of $\mu_{\mathrm{max}}$ photons, it
is sufficient to apply the CAP only to the $\mu_{\mathrm{max}}$th diagonal block, i.e.,
$\left<\Phi_{n,l,m,\mu}\right|W\left|\Phi_{n',l,m,\mu}\right>$ is set to $0$ for 
$\mu \ne \mu_{\mathrm{max}}$.

\section{Calculations}
\label{sec3}

As a test of our method, we determined the one-photon ionization cross section of neutral Xe at a photon
energy of $12.7$ eV. In this calculation, $s$, $p$, and $d$ waves were included, and $\mu_{\mathrm{max}}$
was set to $1$. Let us first consider ionization of the $5p$, $m=0$ level. After assembling the corresponding
real symmetric matrix ${\bm H}_{\mathrm{AT}} + {\bm H}_{\mathrm{EM}} + {\bm H}_{\mathrm{I}}$, those $120$
eigenvectors of this matrix were computed that have the largest overlap with the initial-state vector
$\left|\Phi_{5,1,0,0}\right>$. The $\eta$-dependent complex symmetric eigenvalue problem of ${\bm H}$ 
[Eq. (\ref{eq6})] was then solved in the subspace of the previously calculated eigenvectors of the real part
of ${\bm H}$. The chosen subspace size of $120$ provided converged results and, at the same time, allowed for
extremely fast optimization of the parameter $\eta$ [see Eq. (\ref{eq5}) and the text following it].

The $\eta$ trajectory of the resonance energy in the complex plane, for an intensity of 
$1 \times 10^{11}$ W/cm$^2$, is shown in Fig. \ref{fig2}. On the basis of this graph, a dynamic Stark shift of
$7.31 \times 10^{-6}$ a.u. and an ionization rate of $\gamma_{5p,0} = 3.05 \times 10^{-5}$ a.u. are found.
Proceeding in a similar fashion, the ionization rate of the $5p$, $m=\pm 1$ levels is calculated as 
$\gamma_{5p,\pm 1} = 2.01 \times 10^{-5}$ a.u. at $1 \times 10^{11}$ W/cm$^2$, so that the $m$-averaged 
ionization rate 
\begin{equation}
\label{eq15}
\bar{\gamma}_{5p} = \left(\gamma_{5p,-1} + \gamma_{5p,0} + \gamma_{5p,1}\right)/3
\end{equation}
is $2.35 \times 10^{-5}$ a.u. We then calculate the total ionization rate as
\begin{equation}
\label{eq16}
\Gamma_{5p} = \bar{\gamma}_{5p} \left(6 - q\right) \; ,
\end{equation}
where, for neutral Xe, $q=0$ ($q$ is the atomic charge). This procedure is approximately valid
also for $q>0$, since spin-orbit interaction ensures that the $6 - q$ $5p$ electrons are uniformly
distributed over $m=-1,0,+1$.

We have checked that $\Gamma_{5p}$ is a linear function of the intensity, $I$, in the vicinity of 
$1 \times 10^{11}$ W/cm$^2$. Thus, within the Herman-Skillman-based independent-particle model,
the one-photon ionization cross section of neutral Xe, at a photon energy of $12.7$ eV, is $119$ Mb.
This result, which differs from the experimental cross section \cite{Sams66} by a little more than a
factor of $2$, has been confirmed by us using the same independent-particle model, but treating the 
continuum problem with an R-matrix code \cite{AyGr96} (see also Ref. \cite{MaCo68}).

In order to test whether our CAP-Floquet program is also capable of describing multiphoton physics, 
we investigated the two-photon ionization cross section of neutral Xe at $6.42$ eV. Experimentally,
this is known to be about $4 \times 10^{-50}$ cm$^4$ s \cite{McEd82}. With atomic $s$ through $f$ waves 
and $\mu_{\mathrm{max}}=2$, we calculated a cross section of $6.2 \times 10^{-50}$ cm$^4$ s. Other 
calculations of this quantity, which are similarly accurate, are reported in Refs. \cite{McGu81,GaTa86,HuWe87}.

We calculated the $(q+1)$-photon ionization cross section, $\sigma_{q+1}$, of Xe$^{q+}$ at a photon 
energy of $12.7$ eV (see Table \ref{tab1}) following a strategy analogous to the one described above for 
neutral xenon:
\begin{eqnarray}
\label{eq17}
\sigma_2\left(\mathrm{Xe}^{+}\right) & = & 4.5 \times 10^{-49} \, \mathrm{cm}^{4} \, \mathrm{s} \, , \\
\label{eq18}
\sigma_3\left(\mathrm{Xe}^{++}\right) & = & 4.7 \times 10^{-84} \, \mathrm{cm}^{6} \, \mathrm{s}^2 \, , \\
\label{eq19}
\sigma_4\left(\mathrm{Xe}^{3+}\right) & = & 1.8 \times 10^{-115} \, \mathrm{cm}^{8} \, \mathrm{s}^3 \, , \\
\label{eq20}
\sigma_5\left(\mathrm{Xe}^{4+}\right) & = & 1.1 \times 10^{-148} \, \mathrm{cm}^{10} \, \mathrm{s}^4 \, , \\
\label{eq21}
\sigma_6\left(\mathrm{Xe}^{5+}\right) & = & 2.4 \times 10^{-179} \, \mathrm{cm}^{12} \, \mathrm{s}^5 \, .
\end{eqnarray}

The spatial profile of the VUV-FEL beam in Hamburg, perpendicular to the beam axis, has a 
Gaussian shape \cite{MoWa04}. Let $(\rho,\varphi,z)$ denote cylindrical coordinates with respect to that axis. 
The intensity near the focus (at $z = 0$) may then be written as 
\begin{equation}
\label{eq22}
I(\rho,z,t) = \frac{4 \ln{2}}{\pi \Delta^2(z)} \exp{\left(-\frac{4 \ln{2}}{\Delta^2(z)} \rho^2\right)}
P(t) \; ,
\end{equation}
where
\begin{equation}
\label{eq23}
\Delta(z) = \Delta \sqrt{1 + (z/z_0)^2}
\end{equation}
is the $z$-dependent full-width-at-half-maximum of the Gaussian beam profile. In the experiment described
in Ref. \cite{WaCa04}, $\Delta = 20$ $\mu$m. The beam divergence was $17$ mrad \cite{MoWa04}, from which 
we estimate that $z_0 = 1.2$ mm. The time-dependent pulse power is represented in Eq. (\ref{eq22}) 
by $P(t)$.

Since we are interested in the {\em nonlinear} response of Xe ions to the vuv laser pulses, it is not
permissible to use for $P(t)$ the pulse shape obtained after averaging over many pulses. The temporal 
shape of the individual FEL pulses has not been measured so far, but reliable simulations of the
FEL performance exist \cite{AyBa02,SaSc99}, which are able to reproduce measured FEL parameters and which, 
in addition, provide information about temporal pulse shapes \cite{DoFo04}. Ten representative, simulated
pulses are shown in Fig. \ref{fig3} \cite{Yurk04}. We see that while the averaged pulse may appear
approximately Gaussian (with a pulse width of about $50$ fs), the individual pulses are not. 

Including an attenuation factor of $0.2$ \cite{MoWa04}, which takes into account the finite reflectivity
of the mirrors used to focus the FEL beam into the xenon gas, we solve, for each of the pulses shown in 
Fig. \ref{fig3}, the rate equations
\begin{widetext} 
\begin{eqnarray}
\dot{n}_0(\rho,z,t) & = & -\sigma_1 \frac{I(\rho,z,t)}{\omega} n_0(\rho,z,t) \; , \nonumber \\
\label{eq24}
\dot{n}_1(\rho,z,t) & = & \sigma_1 \frac{I(\rho,z,t)}{\omega} n_0(\rho,z,t) 
-\sigma_2 \left(\frac{I(\rho,z,t)}{\omega}\right)^2 n_1(\rho,z,t) \; , \\
\dot{n}_2(\rho,z,t) & = & \sigma_2 \left(\frac{I(\rho,z,t)}{\omega}\right)^2 n_1(\rho,z,t)
-\sigma_3 \left(\frac{I(\rho,z,t)}{\omega}\right)^3 n_2(\rho,z,t) \; , \nonumber \\
& \vdots & \nonumber
\end{eqnarray} 
\end{widetext}
for the probabilities $n_q(\rho,z,t)$ of finding Xe$^{q+}$ at time $t$ and position $(\rho,z)$ 
[$\varphi$ arbitrary]. (The thermal motion of the ions on a timescale of $100$ fs may, of course, be
neglected.) The initial conditions are $n_0(\rho,z,t\rightarrow -\infty) = 1$ and
$n_q(\rho,z,t\rightarrow -\infty) = 0$ for $q>0$. Let $\kappa$ stand for the gas density in 
the interaction region. Then the total number of Xe$^{q+}$ generated by a given laser pulse reads
\begin{equation}
\label{eq25}
N_q = 2 \pi \kappa \int_{z_{\mathrm{min}}}^{z_{\mathrm{max}}} \mathrm{d}z
\int_0^{\infty} \mathrm{d}\rho \, \rho \, n_q(\rho,z,t\rightarrow +\infty) \; .
\end{equation}
In the laser experiment at DESY, $\kappa = 2.8 \times 10^{13}$ atoms$/$cm$^3$, $z_{\mathrm{min}} = -1$ mm,
and $z_{\mathrm{max}} = +1$ mm \cite{MoWa04}.

We calculated $N_q$ for each of the $10$ laser pulses in Fig. \ref{fig3} and then determined the average
number of Xe$^{q+}$ generated per laser pulse, $<N_q>$, which is depicted in Fig. \ref{fig4}. It is difficult
to assess whether this result can already explain the measurements in Ref. \cite{WaCa04}. According to the
calculation, Xe$^+$ and Xe$^{++}$ dominate by far, which is consistent with the fact that the detector response
to these two ions appeared to be saturated in the experiment \cite{WaCa04}. The mass spectra in Ref. \cite{WaCa04}
have not been calibrated to account for the specific detector response to ions of different charge states
\cite{MoWa04}, so they may not be linearly related to the actual ion production rate. Moreover, several of the 
experimental parameters we employed in our calculation are, in fact, not known very precisely. Among these are the
gas density, $\kappa$, and the spatial beam width, $\Delta$ \cite{MoWa04}. It should be mentioned, in addition, 
that the FEL was not operating under optimal conditions when the data in Ref. \cite{WaCa04} were taken 
\cite{MoWa04}. Therefore, the laser pulses our calculation is based on (Fig. \ref{fig3}) are, on average, more 
intense than in the experiment.

The theoretical model we use also suffers from shortcomings. As mentioned earlier, multiplet splittings
of the valence shell are not considered, which means the intermediate bound states influencing the multiphoton
ionization cross sections may not be sufficiently accurate. The spectral width, $\Delta \omega/\omega$, 
of the laser pulses of, on average, $1$ \% \cite{MoWa04}---which is broader than the Fourier limit of a $50$-fs 
pulse by an order of magnitude---is not included in the present treatment. In the rate equations, Eq. (\ref{eq24}), 
excited-state populations, phase effects, as well as nonsequential multiphoton processes are neglected. The latter, 
however, may be expected to be strongly suppressed in the vuv regime, as confirmed by the measurements in Ref. 
\cite{WaCa04}.

Before concluding, we would like to mention an interesting many-electron effect that leads to an 
enhancement of the multiphoton ionization rate in the xenon ions. In Xe$^+$, it requires $11.3$ eV \cite{Moor71}
to excite one of the two $5s$ electrons to the $5p$ shell. (Within the Herman-Skillman model we find $10.9$ eV.)
One can, therefore, envisage, as one of the paths leading to two-photon ionization of Xe$^+$, the excitation of
a $5s$ electron by a first vuv photon and the subsequent excitation of one of the $5p$ electrons to a virtual
state bound in the channel associated with the $5s$ hole, as illustrated in Fig. \ref{fig5}. Due to electron 
correlation, one of the remaining $5p$ electrons can fill the $5s$ hole, and the excited electron is ejected 
into the continuum. This is a kind of Auger decay of the inner-valence excited ion, resulting in the formation 
of Xe$^{++}$. A similar scenario is conceivable for the more highly charged xenon ions.

The contribution of this to the multiphoton ionization cross section of Xe$^{q+}$ can be crudely estimated as follows.
We choose $\left|\Phi_{5,0,0,0}\right>$ as initial-state vector in our Floquet code, but instead of utilizing a 
CAP, we simply assign an autoionization width $\Gamma_{\mathrm{auto}}$ (i.e. we add 
$-\mathrm{i}\Gamma_{\mathrm{auto}}/2$) to those diagonal elements in the $\mu_{\mathrm{max}}$th diagonal 
block that satisfy $\varepsilon_{n,l} + \varepsilon_{5,1} - \varepsilon_{5,0} > 0$. This condition implies
that the energy released in the $5p \rightarrow 5s$ transition is sufficient to transfer the electron with
quantum numbers $n$ and $l$ into the continuum. A typical lifetime of an inner-shell hole is of the order
of $10$ fs (or shorter), so we set the autoionization width to $65$ meV. 

Let us call $\gamma_{5s}$ the $5s$-mediated ionization rate determined in this way. The total $5s$-mediated 
ionization rate of Xe$^{q+}$ is then, approximately, 
\begin{equation}
\label{eq26}
\Gamma_{5s} = 2 \gamma_{5s} \frac{q}{6} (7-q) \; .
\end{equation}
The factor of $2$ in this expression is needed since there are two $5s$ electrons in all Xe ions considered here.
If we assume that the six $5p$ spin orbitals have equal probability of being occupied, then the probability
that the $5p$, $m=0$ spin orbital with the right spin is unoccupied is $q/6$. After the virtual excitation 
of one of the $5s$ electrons, there are $7-q$ $5p$ electrons available for the absorption of the 
remaining $\mu_{\mathrm{max}}-1$ photons.

We add the cross sections obtained in this way to the respective cross sections in Eqs. (\ref{eq17})-(\ref{eq21}), 
thereby neglecting interference effects. The results are
\begin{eqnarray}
\label{eq27}
\sigma_2\left(\mathrm{Xe}^{+}\right) & = & 4.6 \times 10^{-49} \, \mathrm{cm}^{4} \, \mathrm{s} \, , \\
\label{eq28}
\sigma_3\left(\mathrm{Xe}^{++}\right) & = & 2.0 \times 10^{-82} \, \mathrm{cm}^{6} \, \mathrm{s}^2 \, , \\
\label{eq29}
\sigma_4\left(\mathrm{Xe}^{3+}\right) & = & 3.3 \times 10^{-115} \, \mathrm{cm}^{8} \, \mathrm{s}^3 \, , \\
\label{eq30}
\sigma_5\left(\mathrm{Xe}^{4+}\right) & = & 3.7 \times 10^{-147} \, \mathrm{cm}^{10} \, \mathrm{s}^4 \, , \\
\label{eq31}
\sigma_6\left(\mathrm{Xe}^{5+}\right) & = & 6.4 \times 10^{-179} \, \mathrm{cm}^{12} \, \mathrm{s}^5 \, .
\end{eqnarray}
The quantities most significantly affected by the $5s$-mediated ionization mechanism are the three-photon 
ionization cross section of Xe$^{++}$ and the five-photon ionization cross section of Xe$^{4+}$.

The ionic charge distribution computed using the multiphoton ionization cross sections in 
Eqs. (\ref{eq27})-(\ref{eq31}) is displayed in Fig. \ref{fig6}. Now the production of Xe$^{3+}$ is clearly
visible even on a linear scale: About $3 \times 10^6$ triply charged xenon ions are produced per laser pulse.
The number of Xe$^{6+}$ ions per laser pulse is more than $3000$, which is also not particularly small.

\section{Conclusion}
\label{sec4}

We have investigated in this paper multiphoton ionization of atomic xenon and its ions at a photon energy
of $12.7$ eV, a radiation intensity of order $10^{13}$ W/cm$^2$, and a pulse duration of about $50$ fs.
A recent experiment employing the VUV-FEL at DESY has demonstrated, under these laser conditions, the 
generation of xenon charge states of up to $6+$ \cite{WaCa04}. In the infrared, even a pulse that is three
orders of magnitude longer (with about the same intensity), produces charge states no higher than Xe$^{4+}$
\cite{HuLI83}. 

Using an effective one-particle model, in combination with the Floquet concept and a complex absorbing potential,
we have calculated vuv multiphoton ionization cross sections that refer to the absorption, by a $5p$ electron,
of as many photons as are needed to ionize it. We have also estimated the influence of $5s$ excitation on the 
ionization cross sections and found that it may be significant. Although we grant that the model we applied is 
not ideal for describing many-electron phenomena (a better many-body calculation would be desirable), the result 
of our estimate indicates that at vuv photon energies multiphoton ionization is driven, at least partly, by 
electronic many-body physics. Focussing on the behavior of a single active electron does not appear to be sufficient.

Taking a rate-equation approach and utilizing simulated FEL pulses \cite{Yurk04}, we determined the average
number of Xe$^{q+}$ ions produced per vuv laser pulse. This step depends heavily on a number of important
experimental parameters \cite{MoWa04}. In our calculation, thousands of Xe$^{6+}$ ions are found to be generated 
per pulse, and a correspondingly higher number for the lower charge states. Hence, the experimental observation of
multiple ionization of xenon, Ref. \cite{WaCa04}, appears compatible with the nonlinear absorption
of several photons. We conclude that multiphoton physics is indeed relevant for some processes driven by the 
intense vuv beam of the free-electron laser in Hamburg. For more quantitative comparisons, it will be desirable
for future experiments to obtain a calibrated ionic charge distribution, as well as more detailed information
about the FEL pulse properties.

\begin{figure}[t]
\includegraphics[width=9.5cm,origin=c,angle=0]{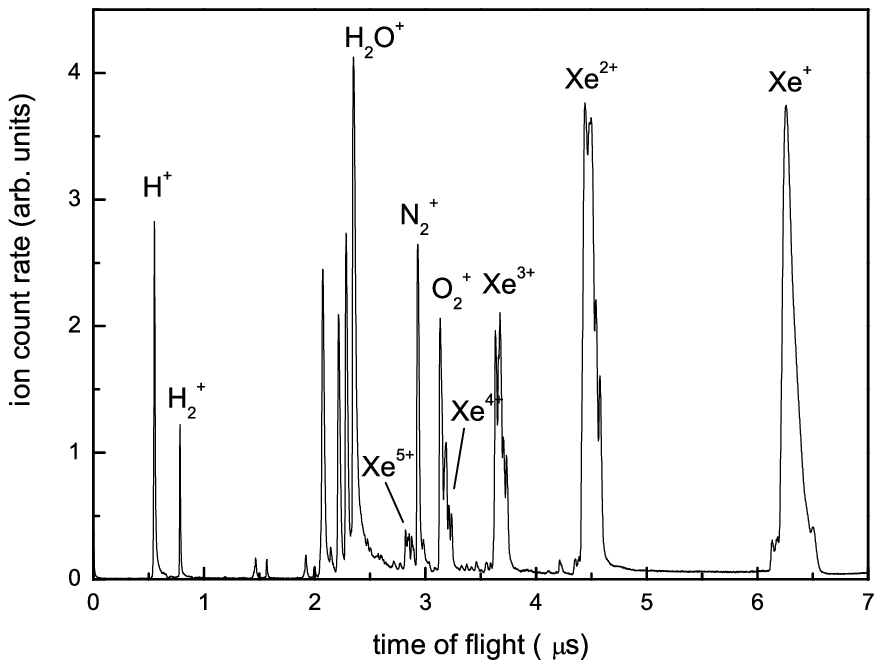}
\caption[]{Experimental \cite{MoWa04} ion detection signal is plotted versus the ionic
time-of-flight. Free-electron laser pulses with a peak intensity of about $10^{13}$ W/cm$^2$, a duration of 
approximately $100$ fs, and a photon energy of $12.7$ eV produce, in an atomic xenon beam, ionic species of 
various charge states \cite{WaCa04}. This time-of-flight mass spectrum \cite{MoWa04} is 
an average over $100$ consecutive FEL pulses.}
\label{fig1}
\end{figure}

\begin{figure}[t]
\includegraphics[width=9.5cm,origin=c,angle=0]{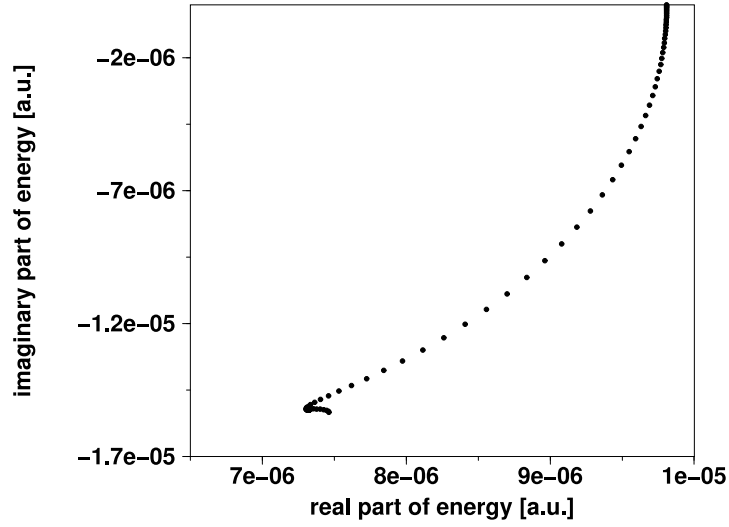}
\caption[]{The dots in this figure correspond to a specific eigenvalue of the family of complex
symmetric matrices ${\bm H}(\eta_n)$ [Eq. (\ref{eq6})], where $\eta_n = \delta (\kappa^n - 1)/(\kappa - 1)$
[$n=0,\ldots,99$; $\delta = 3\times 10^{-9}$; $\kappa=1.1$]. In the absence of photons and absorbing potential,
this eigenvalue equals the $5p$, $m=0$ level of atomic xenon (which defines the origin in the figure). Due to the
interaction with $12.7$-eV photons, this level is shifted as well as broadened. The point of stabilization
of the $\eta$ trajectory implies a dynamic Stark shift of $7.31 \times 10^{-6}$ a.u. and an ionization rate
of $3.05 \times 10^{-5}$ a.u., at a radiation intensity of $1 \times 10^{11}$ W/cm$^2$.}
\label{fig2}
\end{figure}

\begin{figure}[t]
\includegraphics[width=9.5cm,origin=c,angle=0]{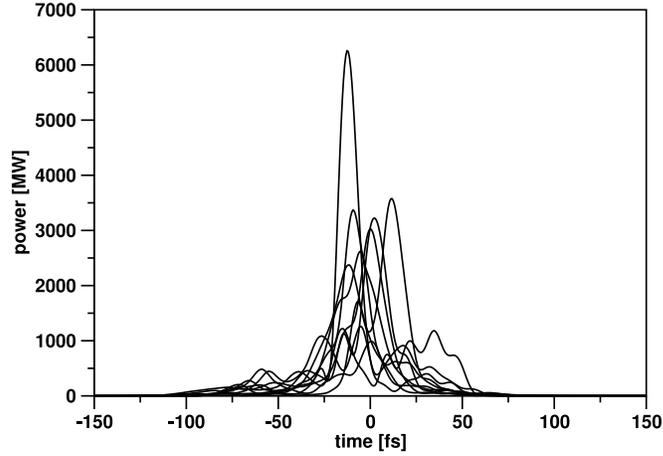}
\caption[]{The power in the DESY free-electron laser pulse is shown as a function of time.
The vuv laser pulses supplied by the free-electron laser source at DESY, Hamburg, are ultrashort,
with an average width of about $50$ fs, and intense, with a pulse energy of order $10$ $\mu$J \cite{AyBa02}. 
The pulses shown in this figure \cite{Yurk04} have been calculated using an FEL simulation program 
\cite{SaSc99,DoFo04}.}
\label{fig3}
\end{figure}

\begin{figure}[t]
\includegraphics[width=9.5cm,origin=c,angle=0]{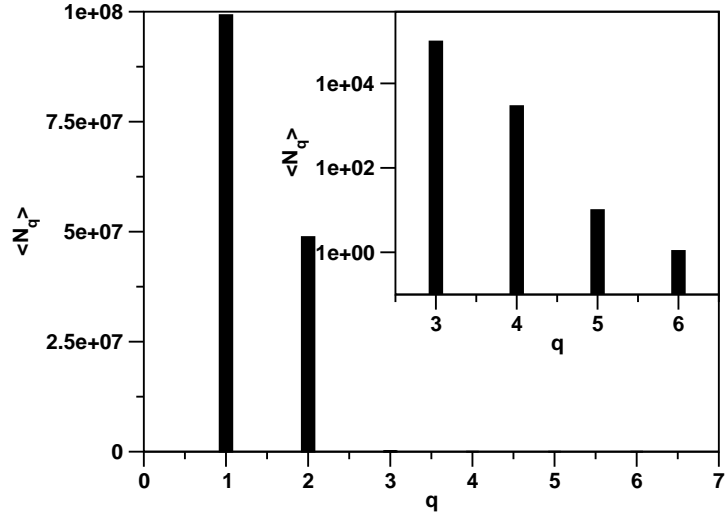}
\caption[]{The average number, $<N_q>$, of Xe$^{q+}$ ions produced per VUV-FEL laser pulse, as calculated 
on the basis of the multiphoton ionization cross sections in Eqs. (\ref{eq17})-(\ref{eq21}), the laser pulse
properties [Eqs. (\ref{eq22}), (\ref{eq23}) and Fig. \ref{fig3}], the rate equations in Eq. (\ref{eq24}), 
and the integral over the interaction volume in Eq. (\ref{eq25}). Note the logarithmic scale along the 
ordinate of the inset.}
\label{fig4}
\end{figure}

\begin{figure}[t]
\includegraphics[width=9.5cm,origin=c,angle=0]{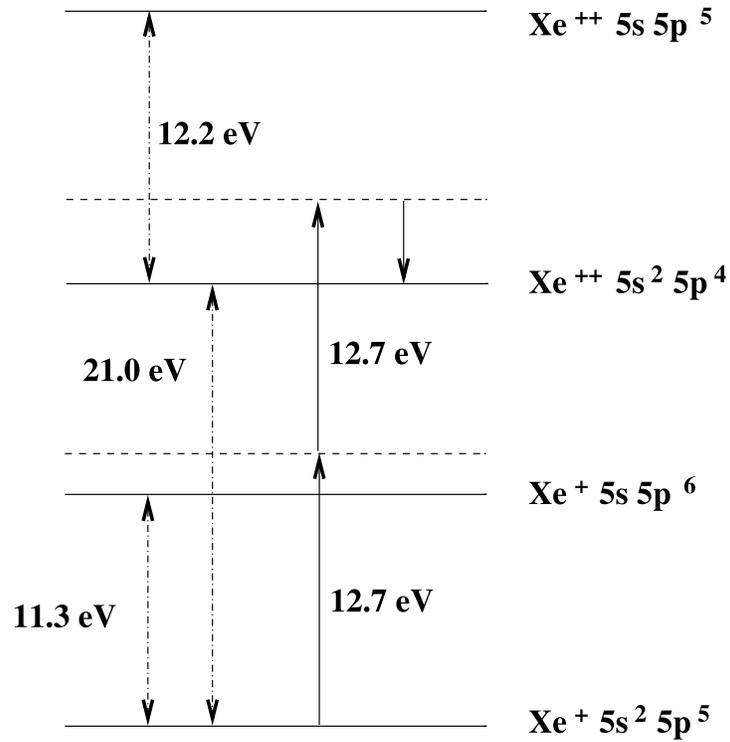}
\caption[]{Schematic depiction of energy levels associated with the excitation of xenon ions from the
$5s$ subshell. For instance, Xe$^+$ can be photoionized via the absorption of two vuv photons by a $5p$ electron. 
A second ionization path, alluded to in this figure, involves a photon absorption that virtually excites a $5s$ 
electron to the $5p$ shell. Simultaneously, a $5p$ electron is excited to an autoionizing state associated with a Rydberg
series converging to the $5s$~$5p^5$ threshold of Xe$^{++}$. Electron correlation then induces a transition
from $5p$ to $5s$ accompanied by the emission of an electron.}
\label{fig5}
\end{figure}

\begin{figure}[t]
\includegraphics[width=9.5cm,origin=c,angle=0]{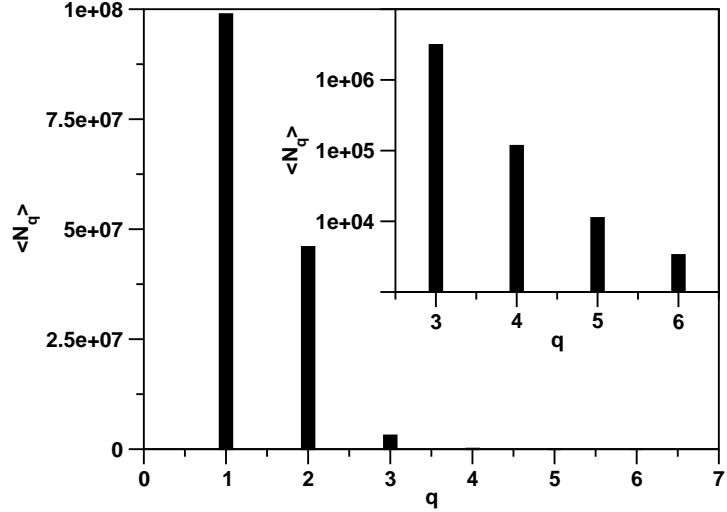}
\caption[]{The same as in Fig. \ref{fig4}, the only difference being the fact that the multiphoton ionization
cross sections underlying the calculation were taken from Eqs. (\ref{eq27})-(\ref{eq31}), not 
(\ref{eq17})-(\ref{eq21}). $<N_q>$ is the average number of Xe$^{q+}$ ions produced per VUV-FEL radiation pulse.}
\label{fig6}
\end{figure}

\acknowledgments
We would like to thank Thomas M\"oller and Hubertus Wabnitz for providing us with valuable
information about their experiment and for contributing Fig. \ref{fig1}. We also wish to express 
our gratitude to Mikhail Yurkov for supplying his calculated FEL pulse data. Financial support by the 
U.S. Department of Energy, Office of Science is gratefully acknowledged.

\pagebreak

\pagebreak

\pagebreak

\pagebreak

\pagebreak

\pagebreak

\pagebreak

\begin{table}[t]
\caption[]{\label{tab1} X$\alpha$ parameters employed to reproduce the $5p$ ionization potential 
(I.P.) of Xe$^{q+}$, i.e. the energy needed to remove one electron from Xe$^{q+}$ and generate 
Xe$^{(q+1)+}$ in its ground state. Also shown is the minimum number of $12.7$-eV photons (N.P.)
needed to ionize Xe$^{q+}$.}
\begin{ruledtabular}
\begin{tabular}{cccc}
$q$ & I.P. [eV] & X$\alpha$ & N.P.\\
\hline
$0$ & $12.1$\footnotemark[1] & $1.067$ & $1$ \\
$1$ & $21.0$\footnotemark[2] & $1.031$ & $2$ \\
$2$ & $33.1$\footnotemark[3] & $1.180$ & $3$ \\
$3$ & $42$\footnotemark[4]   & $1.056$ & $4$ \\
$4$ & $53$\footnotemark[5]   & $1.044$ & $5$ \\
$5$ & $64$\footnotemark[6]   & $0.999$ & $6$ \\
\end{tabular}
\end{ruledtabular}
\footnotetext[1]{Ref. \cite{BrVe01}}
\footnotetext[2]{Ref. \cite{HaPe87}}
\footnotetext[3]{Ref. \cite{MaBa87}}
\footnotetext[4]{Ref. \cite{GrDi83}}
\footnotetext[5]{Ref. \cite{DoMo59}}
\footnotetext[6]{Ref. \cite{Syag92}}
\end{table}

\end{document}